\documentclass{article}
\usepackage{cite}
\usepackage[utf8]{inputenc}
\usepackage{amsmath}
\usepackage{amssymb,amsfonts}
\usepackage{algorithmic}
\usepackage{graphicx}
\usepackage{textcomp}
\usepackage{xcolor}
\usepackage{caption} 
\usepackage{subcaption}
\usepackage{float}
\usepackage{siunitx}
\DeclareUnicodeCharacter{0394}{$\Delta$}
\DeclareUnicodeCharacter{2009}{\,} 
\def\BibTeX{{\rm B\kern-.05em{\sc i\kern-.025em b}\kern-.08em
    T\kern-.1667em\lower.7ex\hbox{E}\kern-.125emX}}
\graphicspath{ {images/} }
\begin{document}

\title{Screening of potential double perovskite materials for photovoltaic applications using agglomerative hierarchical clustering}

\author{Utkarsh Saha\textsuperscript{1}, Koyendrila Debnath\textsuperscript{2}, Soumitra Satapathi\textsuperscript{1{*}}}

\maketitle

\noindent1. Department of Physics, Indian Institute of Technology Roorkee, Roorkee, India \\
2. Theoretical Sciences Unit, Jawaharlal Nehru Centre for Advanced Scientific Research, Bangalore, India \\
{*}corresponding author:
Soumitra Satapathi (soumitrasatapathi@gmail.com)

\begin{abstract}
Data-driven approaches to solve problems in materials science have gained immense popularity in recent times due to their ability to predict unknown material properties and uncover relationships between structure and property. Machine learning algorithms like GBRT, random forest and neural networks have had tremendous success in predicting target properties of materials and design of structures for various applications. However, a major drawback for achieving results within the required accuracy using these algorithms has been the need for large datasets which can be challenging for problems when data is not sufficiently available for training the models. In this work, we propose the use of a hierarchical clustering algorithm which can work considerably better on materials science problems with small dataset constraints. We apply the algorithm to screen out promising double perovskite materials as candidates for solar cells. 
\end{abstract}

\section*{Introduction}

In recent years, there has been growing interest in inorganic double perovskites for applications in solar cells and light-emitting diodes because of adjustable photoelectric properties.  \cite{zhao2020facile, li2020novel} As hybrid organic-inorganic perovskites (HOIPs) \cite{li2016recent} suffer from instability and toxicity issues and ABX$_{3}$ perovskites generally have wide band gaps not suitable for photovoltaic applications, replacing A-site or B-site cations of perovskites with two cations forming double perovskites \cite{zhang2020halide, zhao2018rational} may open the path for a more stable class of perovskites. This could lead us to a novel category of materials capable of achieving both the superior performance of HOIPs and the thermal stability of ABX$_{3}$ inorganic perovskite materials. However, there is limited data on double perovskites \cite{wang2020photoactive, wang2020lead} which makes it challenging for researchers to explore their extensive properties. Therefore, exploring high-performance double perovskites in photovoltaics has huge research and development prospects.\\

Searching for optimal materials for any application is a time-consuming process.  Generally, for the discovery of new materials, the first process involves the selection of materials followed by the prediction of material properties. If satisfactory results are achieved, chemical synthesis can be targeted and experimental validation is repeated until adequate performance is achieved. This experimental approach lacks the expeditious nature of materials discovery that we desire in order to focus on growing challenges in the domain of sustainable energy materials. Other predictive techniques have gained popularity in recent years, for example, high-throughput computational screening (HTCS) which involves extensive property prediction using DFT calculations. For this approach, screening of materials in a suitable search space is first carried out using molecular libraries and open databases such as Materials Project \cite{jain2013commentary}, ICSD \cite{bergerhoff1987crystallographic}, Cambridge Structural Database \cite{groom2016cambridge} and various other repositories. Potential candidates for various applications are efficiently categorized using this technique and then they are sent for subsequent experimental verification. This procedure has been progressively used in materials science for various applications such as crystal structure prediction \cite{ryan2018crystal}, using only elemental composition for learning the chemistry of materials \cite{jha2018elemnet}, and target property prediction \cite{xie2018crystal, ye2018deep}. However, the main drawback of HTCS is that the computational complexity increases with the size of the molecule or compound and we have to compromise on faster computational results if we are targeting the results up to a certain accuracy. DFT calculations are also computationally quite expensive and need high-performance computing for implementing advanced DFT methods. \\

To overcome the challenges associated with HTCS, machine learning algorithms have emerged as one of the most efficient ways for the methodical discovery of novel materials. Machine learning algorithms have been used to solve problems in materials for energy storage \cite{chen2020machine}, optoelectronics \cite{lu2020deep}, polymers \cite{wu2019machine}, thermoelectrics \cite{iwasaki2019machine} and a variety of other fields. Data-driven approaches for the design of materials have accelerated the efficiency of material search in a vast computational space. The hidden complexity of physical systems is also uncovered using ML algorithms. A variety of ML algorithms based on gradient boosted regression trees \cite{im2019identifying}, deep generative models \cite{van2020brain, chen2021learning}, support vector machines \cite{wu2020machine, moosavi2020role} and random forest algorithms \cite{torrisi2020random, wu2020machine} have been implemented to achieve results with remarkable accuracy. In the realm of photovoltaics especially, machine learning has opened new doors for the discovery of novel perovskite materials \cite{tao2021machine}. An ML algorithm based on elemental descriptors which was able to predict the Eg values of AA’BB’O$_{6}$ double perovskites was developed in 2016 by Pilania et al. \cite{pilania2016machine}. Xu et al. \cite{xu2018rationalizing} discovered a strategy to identify the formability of all ABX$_{3}$ and AA'BB'X$_{6}$ compounds stored in the Materials Projects database. In 2019, Agiorgousis et al. \cite{l2019machine} used a random forest algorithm to investigate chalcogenide double perovskites in order to find photovoltaic absorbers that can replace CH$_{3}$NH$_{3}$PbI$_{3}$. However, most of these works rely on supervised learning algorithms which make use of large datasets (greater than 1000 compounds) to achieve accuracies of greater than 90\%. In many materials science problems, the availability of data is scarce and building such large datasets to train our model is not feasible.\\

To overcome the above-mentioned issue, unsupervised learning algorithms have been developed in recent years for predicting potential materials with specific target properties for an application. There are a few unsupervised learning approaches that have been implemented in the domain of materials science over the past few years. Amanda et al. have explored clustering methods \cite{parker2019selecting} and Goldsmith et al. have made use of subgroup discovery to uncover structure-property relationships \cite{goldsmith2017uncovering}. Recently, Tao et al. executed an unsupervised learning algorithm for thin-film materials discovery in photovoltaics \cite{wang2021unsupervised}. Clustering is one of the prominent algorithms in unsupervised ML and it plays a key role in categorizing large amount of information into a few number of clusters from which we can extract some meaningful information \cite{patel2015study, karthikeyan2011hierarchical, hourdakis2010hierarchical}. \\

In our study, we propose an agglomerative hierarchical clustering algorithm and apply it to a dataset comprising 540 halide double perovskites of the form A$_{2}$BB’X$_{6}$. The compounds are selected from two space groups, cubic and orthorhombic and a number of features are considered for each compound to obtain a 32-dimensional feature vector for each compound which is then fed as input to the algorithm. We make use of a partition such that we get 10 clusters of compounds. Then, we check which of these clusters has the highest percentage of compounds in the band gap range 1.1 - 1.8 eV by making use of band gap data available in open repositories like the Materials Project for our initial screening. Then, we select the unknown compounds in the group and send them for further DFT studies. The workflow of the entire process is shown in Fig. 1. Our calculations show that 8 of these compounds have properties that could be useful for photovoltaic applications.

\begin{figure}[H]
\centering
\includegraphics[width=0.8\textwidth]{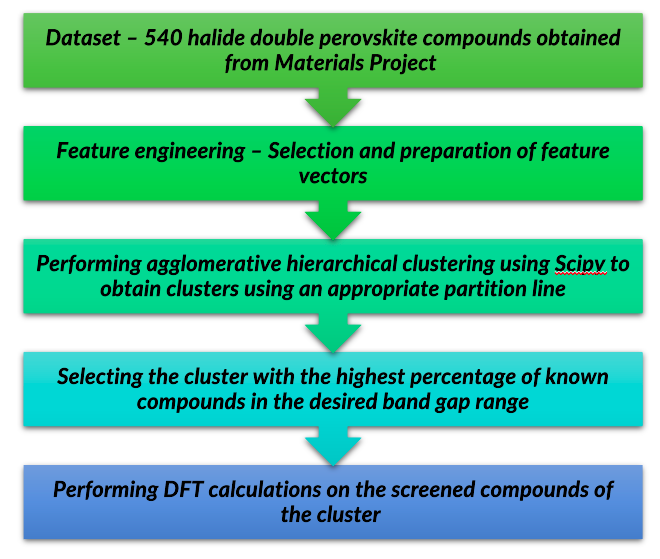}
\caption{Workflow for the guided discovery of halide double perovskites using AHC algorithm}
\label{}
\end{figure}

\section*{Results}

\subsection*{Data Collection}

For preparing the halide double perovskite dataset, we considered a series of alkali metals for the A-site, K, Rb and Cs. For the B site, we took into account the elements Cu, Ag and Au from Group XI and Tl from Group XIII. For the B’ site, Group XIII elements like Al, Ga, In and lower group XV elements As, Sb, Bi were considered. Cl, Br and I were assigned as the halide anions. For simplicity of calculations, organic molecules such as methyl ammonium for the A-site were not included. Fig. 2(a) shows the crystal structure of the double perovskites and Fig. 2(b) shows the compositional space of the double perovskites spanned by various elements from the periodic table. A total of 270 compounds were formed as a result of the various combinations of the atoms on the four sites. Considering two space groups, cubic and orthorhombic, for each compound, we obtain a total of 540 entries in our dataset. The band gap data and heat of formation data needed for checking the results of the clustered compounds in further stages were collected using open repositories such as the Materials Project and ICSD for each compound.

\begin{figure}[H]
\centering
\begin{subfigure}{.8\textwidth}
    \centering
    \includegraphics[width=.5\textwidth]{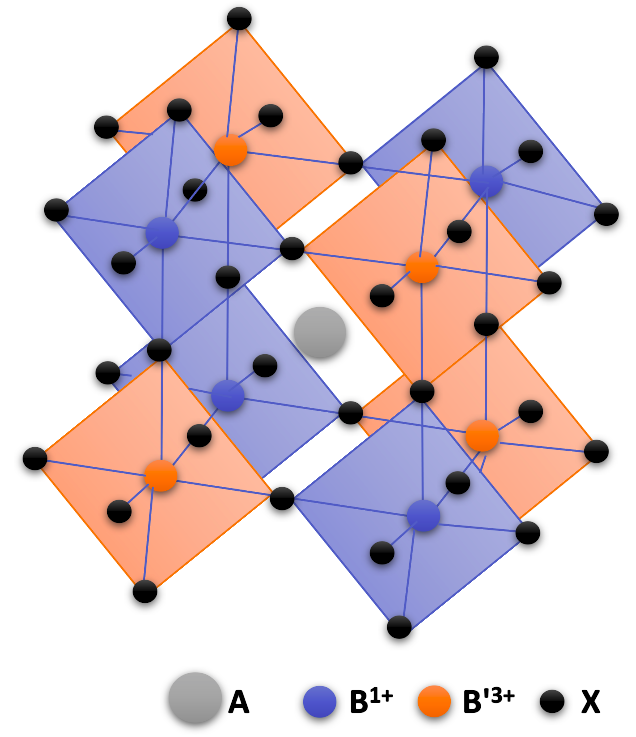}
    \label{}
\end{subfigure}
\begin{subfigure}{1.2\textwidth}
    \centering
    \hspace*{-0.5cm}\includegraphics[width=.9\textwidth]{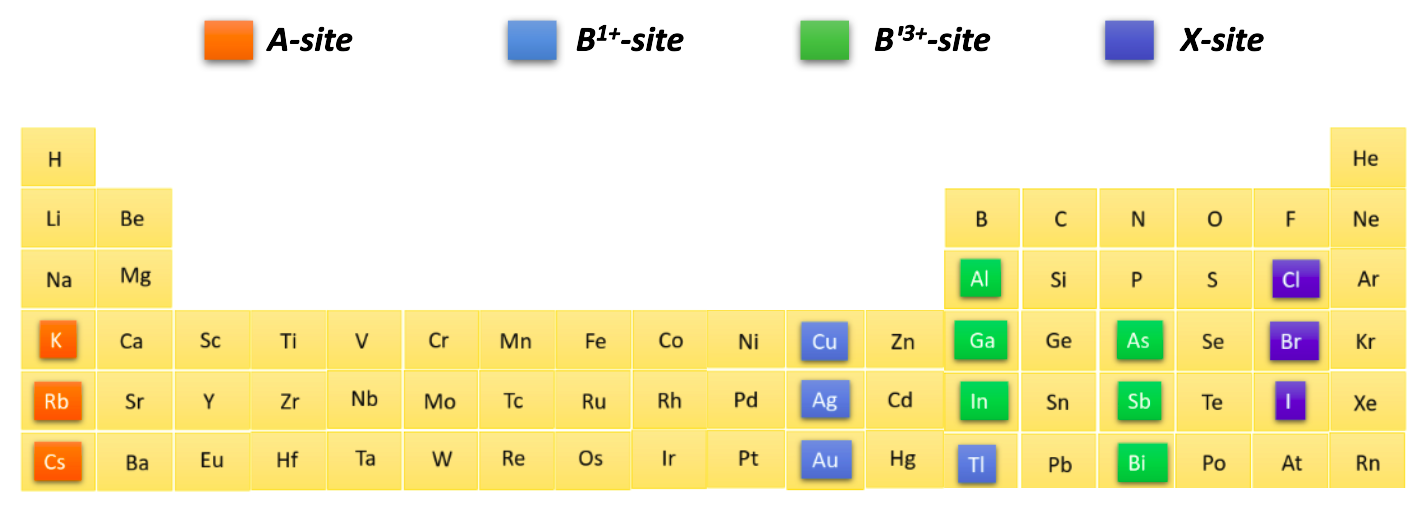}\hspace*{1.5cm}
    \label{}
\end{subfigure}
\caption{Crystal structure of halide double perovskite with A, B$^{1+}$, B$^{3+}$, X-sites, and Chemical elements considered in the dataset}
\label{}
\end{figure}

\subsection*{Preparing feature vectors}

We first consider the band gap as our target property as it is a fundamental parameter for studying the electronic properties of optoelectronic materials. Hence, for our initial task, we need to develop a feature set in order to create a mapping between the compounds and their electronic properties. More number of features usually enables our model to perform better but our feature set has to be limited. Previous works \cite{lu2018accelerated, zhuo2018predicting, gladkikh2020machine, wang2021accelerated} show that the elemental properties of materials have strong correlations with their band gaps. Therefore, we select eight elemental properties for each compound and include the space group as an additional feature (0 for cubic and 1 for orthorhombic). These elemental properties comprise ionization energies, electronegativities, highest occupied and lowest unoccupied atomic levels, etc. The complete tabulated list of all the properties is provided in Supplementary Table 1.  A total of 34 features combined from the four elements were listed and a 34-feature vector was created for each compound. We then proceed to the next step which is applying the algorithm.

\subsection*{Applying the AHC algorithm}

The 540 x 34 matrix was inserted as input for the agglomerative hierarchical clustering algorithm to cluster the 540 A$_{2}$BB’X$_{6}$ compounds based on their features. We made use of a partition line to group the compounds into 10 clusters, ranging from cluster 1 to 10 (C1, C2, ... , C10) as shown in Fig. 3. \\ 
\begin{figure}[H]
\centering
\includegraphics[width=1.0\textwidth]{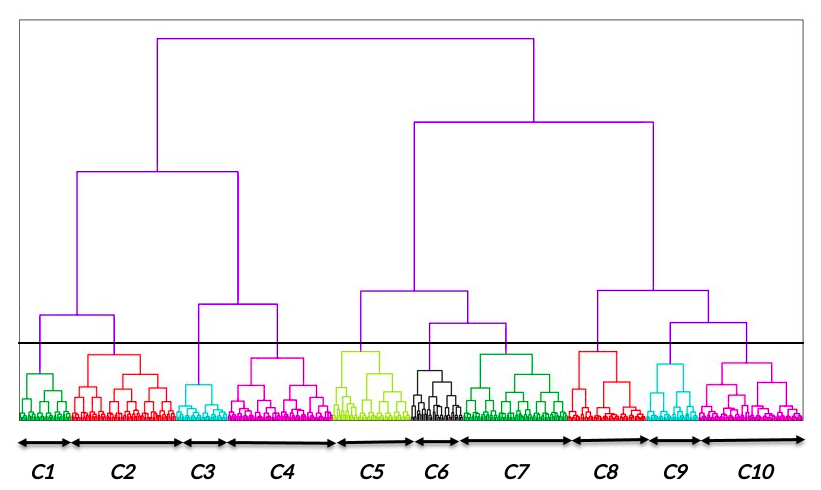}
\caption{Dendrogram plot generated by agglomerative hierarchical clustering}
\label{}
\end{figure}

The features share similar characteristics within the same cluster and the clusters are well-differentiated (C1, C3, C6, C9 having 36 compounds each, C2, C4, C7, C10 having 72 compounds each and C5, C8 having 54 compounds each) as shown in Fig. 4. From the second plot data in Fig. 4, we see that most of the known compounds with band gaps in the range 1.1 - 1.8 eV (compared with band gap data from Materials Project \& ICSD) are clustered in C7, accounting for more than 50\% of the total compounds. \\

\begin{figure}[hbt!]
\centering
\begin{subfigure}{1.8\textwidth}
    \centering
    \hspace*{-5cm}\includegraphics[width=.5\textwidth]{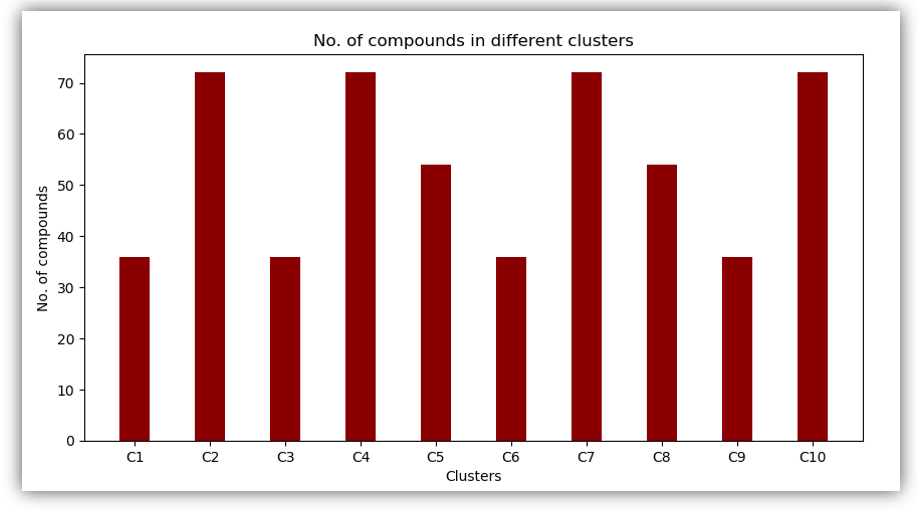}\hspace*{5cm}
    \label{}
\end{subfigure}
\begin{subfigure}{0.8\textwidth}
    \centering
    \includegraphics[width=.9\textwidth]{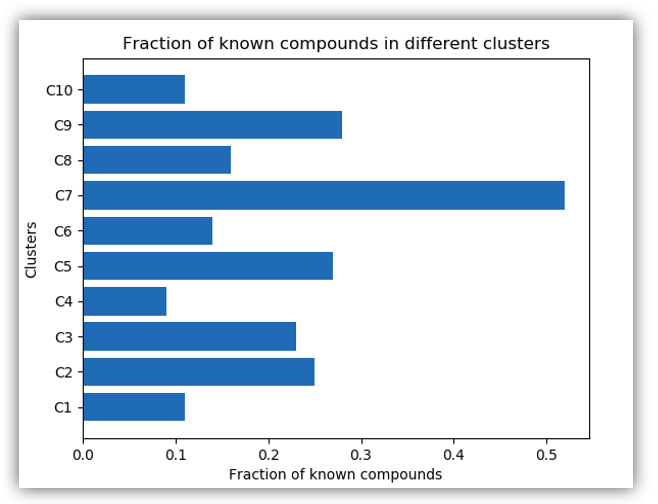}
    \label{}
\end{subfigure}
\caption{No. of componds clustered in each of the 10 clusters, and the percentage of known compounds in the band gap range 1.1-1.8 eV}
\label{}
\end{figure}

A few of these structures include Cs$_{2}$AgBiBr$_{6}$ (E$_{g}$ = 1.622 eV),  Rb$_{2}$AgBiBr$_{6}$ (E$_{g}$ = 1.597 eV), Cs$_{2}$AgAsCl$_{6}$ (E$_{g}$ = 1.606 eV), etc. The next highest percentage of known compounds in the desired band gap range are found in C9 (27\%), C5 (26\%), C2 (24\%) and C3 (22\%). The remaining clusters have less than 15\% known compounds required. Observing such a large margin of difference between C7 and the other clusters led us to investigate the compounds of C7 further and look at the compounds present which have not been explored properly or reported yet in the literature. We were able to pick out 18 unexplored compounds out of the 72 clustered compounds. Then, structural tolerance factor was used as a criterion to further eliminate compounds from the 18 compounds which did not meet the required criterion. We study the remaining 10 compounds further using DFT calculations in the next step.

\newpage

\subsection*{Electronic structure calculations}

After our final screening, we try to calculate the band gaps of the remaining compounds of C7 on which DFT studies have not been performed yet. As per our knowledge, data regarding these compounds is only available in open databases and a thorough study of their electronic structure hasn't been carried out yet. We use VASP for the band gap calculations using the PAW pseudopotentials and
\begin{figure}[hbt!]
\centering
\begin{subfigure}{.5\textwidth}
  \centering
  \includegraphics[width=.5\linewidth]{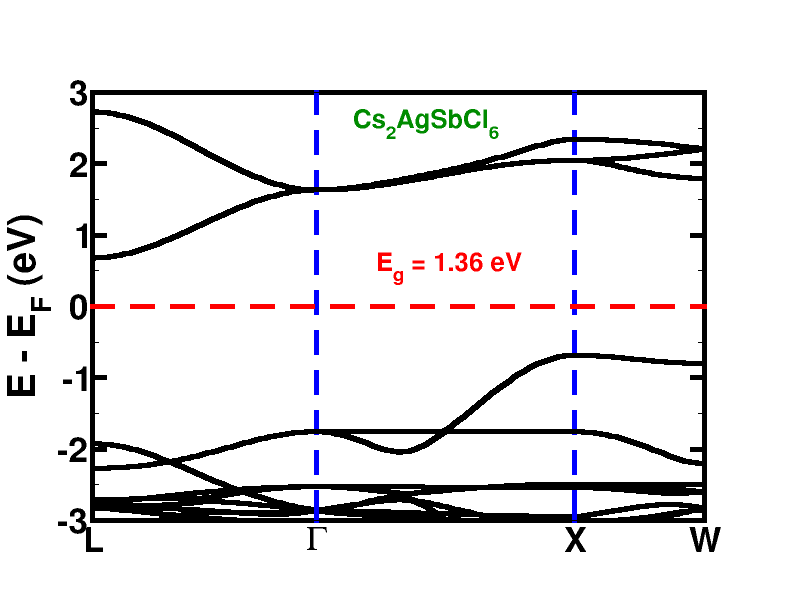}
  \caption{Cs$_{2}$AgSbCl$_{6}$}
  \label{}
\end{subfigure}%
\begin{subfigure}{.5\textwidth}
  \centering
  \includegraphics[width=.5\linewidth]{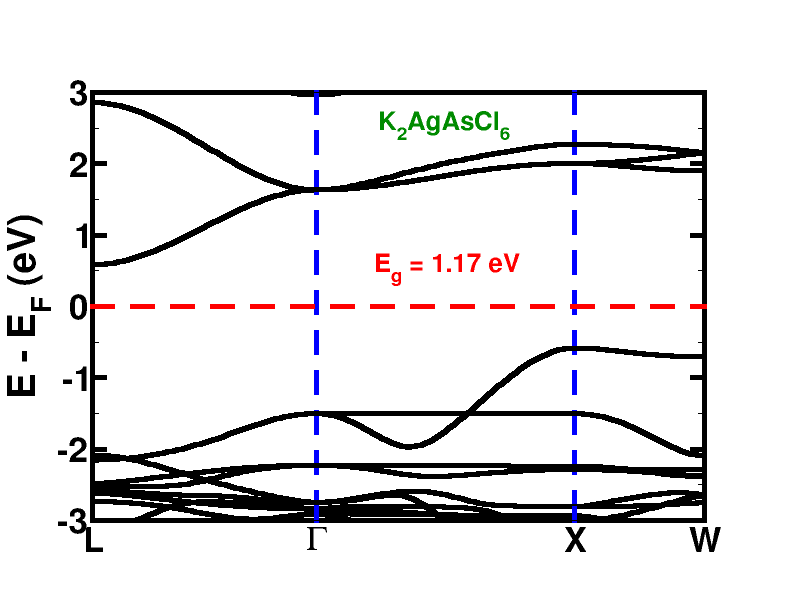}
  \caption{K$_{2}$AgAsCl$_{6}$}
  \label{}
\end{subfigure}
\begin{subfigure}{.5\textwidth}
  \centering
  \includegraphics[width=.5\linewidth]{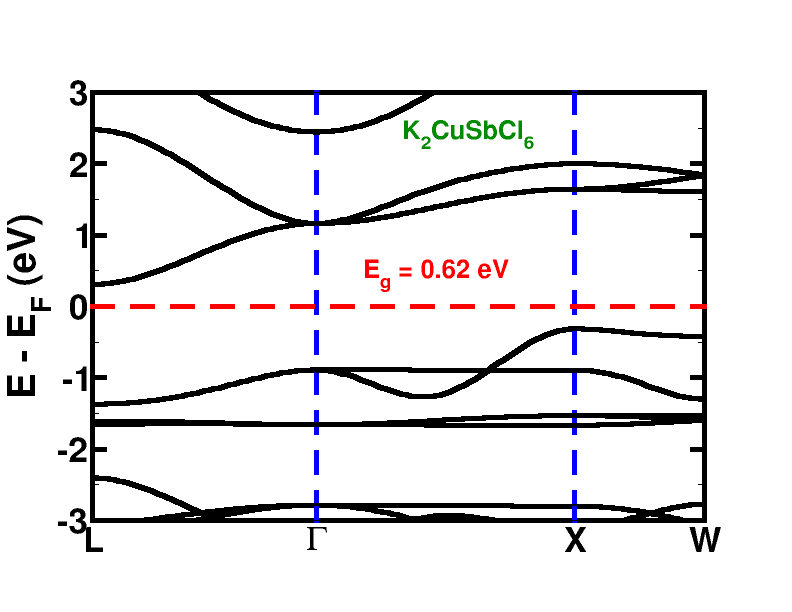}
  \caption{K$_{2}$CuSbCl$_{6}$}
  \label{}
\end{subfigure}%
\begin{subfigure}{.5\textwidth}
  \centering
  \includegraphics[width=.5\linewidth]{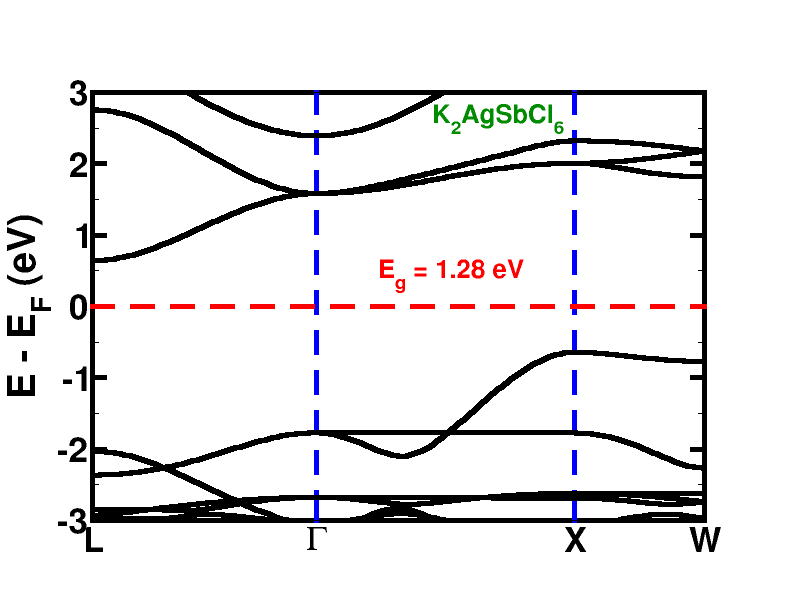}
  \caption{K$_{2}$AgSbCl$_{6}$}
  \label{}
\end{subfigure}
\begin{subfigure}{.5\textwidth}
  \centering
  \includegraphics[width=.5\linewidth]{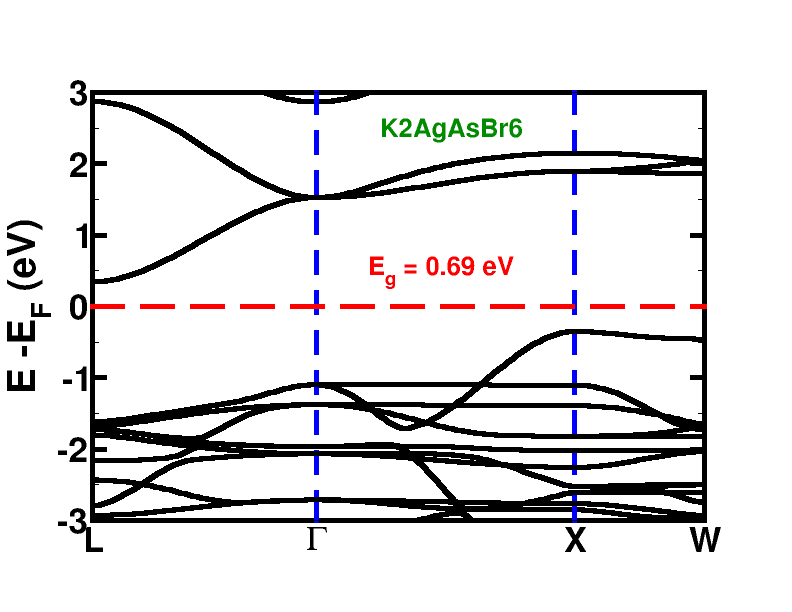}
  \caption{K$_{2}$AgAsBr$_{6}$}
  \label{}
\end{subfigure}%
\begin{subfigure}{.5\textwidth}
  \centering
  \includegraphics[width=.5\linewidth]{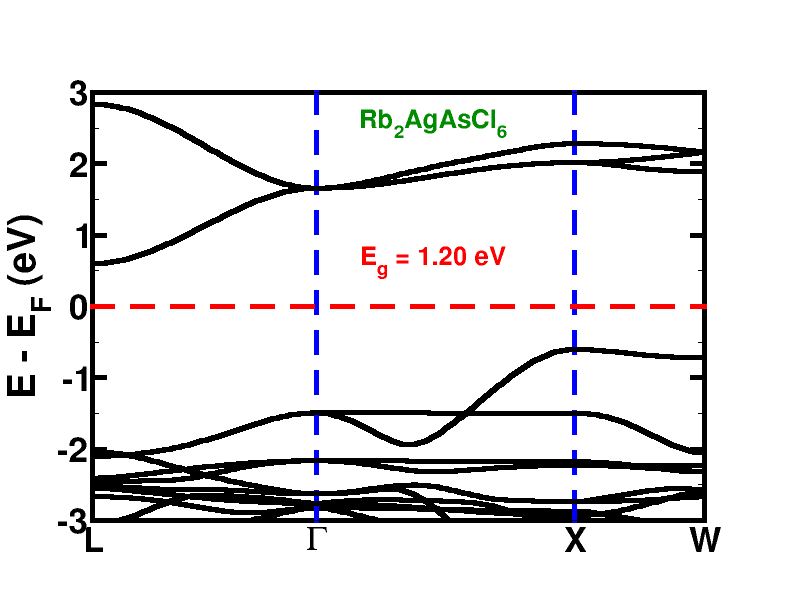}
  \caption{Rb$_{2}$AgAsCl$_{6}$}
  \label{}
\end{subfigure}
\begin{subfigure}{.5\textwidth}
  \centering
  \includegraphics[width=.5\linewidth]{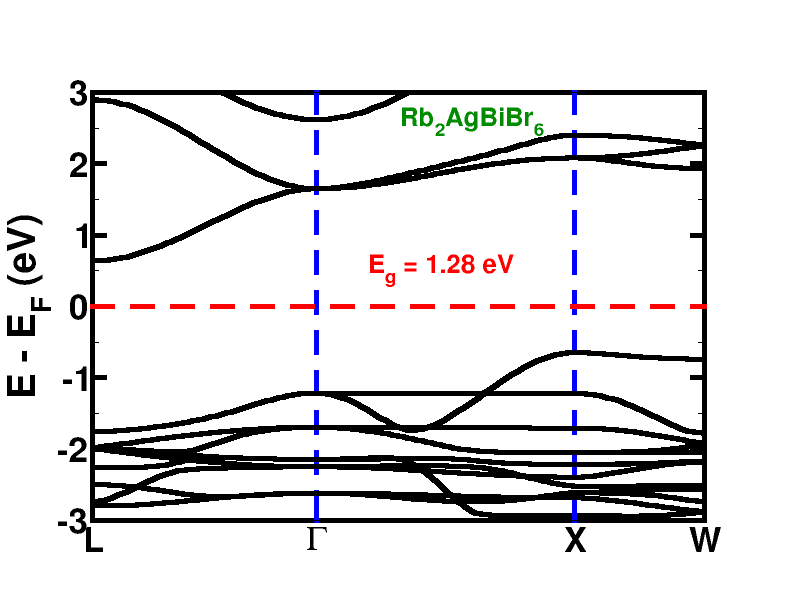}
  \caption{Rb$_{2}$AgBiBr$_{6}$}
  \label{}
\end{subfigure}%
\begin{subfigure}{.5\textwidth}
  \centering
  \includegraphics[width=.5\linewidth]{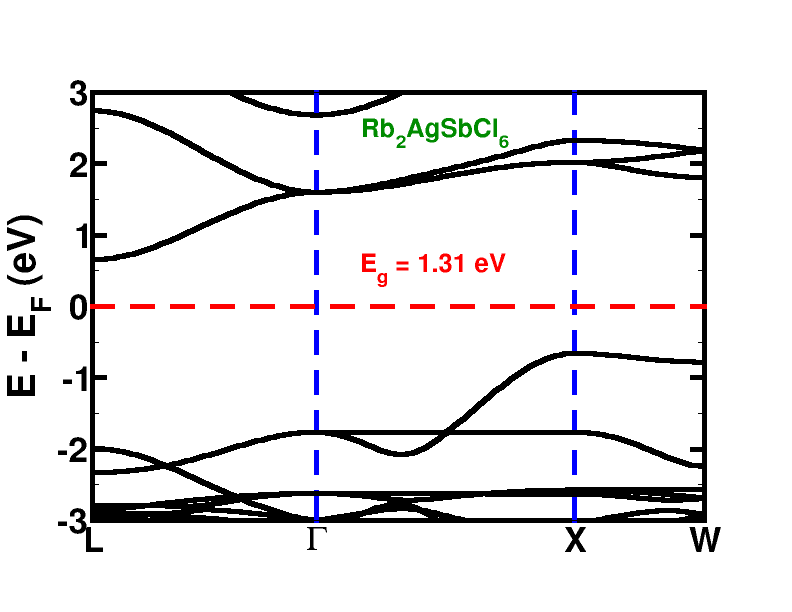}
  \caption{Rb$_{2}$AgSbCl$_{6}$}
  \label{}
\end{subfigure}
\begin{subfigure}{.5\textwidth}
  \centering
  \includegraphics[width=.5\linewidth]{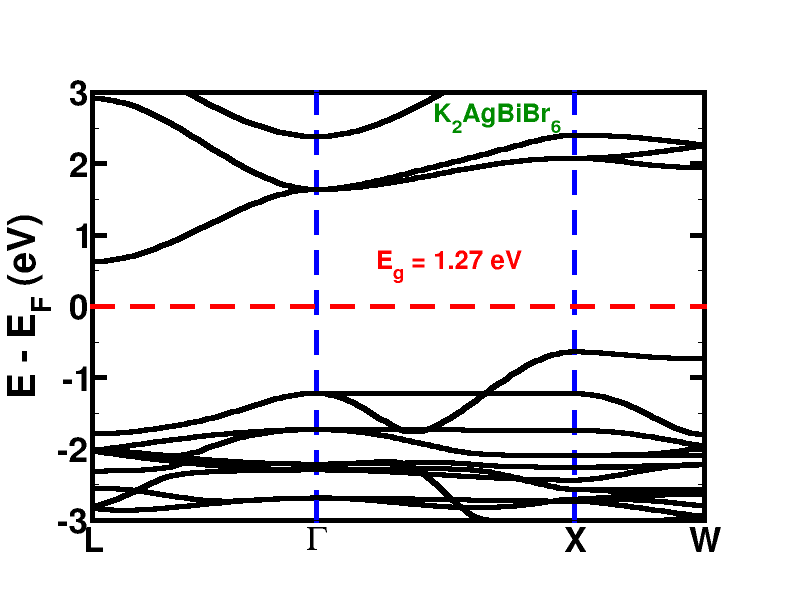}
  \caption{K$_{2}$AgBiBr$_{6}$}
  \label{}
\end{subfigure}%
\begin{subfigure}{.5\textwidth}
  \centering
  \includegraphics[width=.5\linewidth]{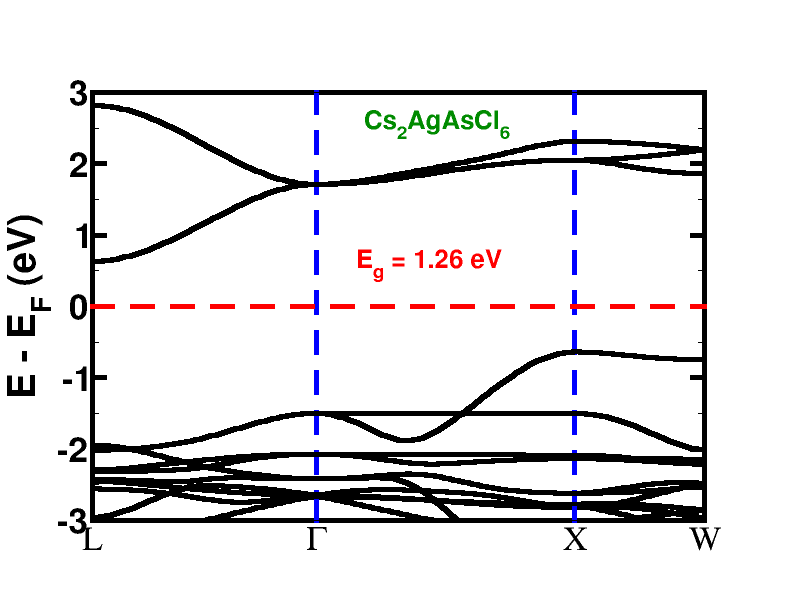}
  \caption{Cs$_{2}$AgAsCl$_{6}$}
  \label{}
\end{subfigure}

\caption{Bandstructure plots}
\label{}
\end{figure}
 PBE exchange correlation. The Monkhorst-Pack k-point mesh of 12 x 12 x 12 is employed. The hybrid functional HSE06 is used for the preliminary studies, along with the SOC effect. The bandstructure plots for each compound are plotted in Fig. 5. \\ 
 From the bandstructure plots, we see that 8 of the 10 compounds have their band gaps in our desired band gap range (Cs$_{2}$AgSbCl$_{6}$, K$_{2}$AgAsCl$_{6}$, K$_{2}$AgSbCl$_{6}$, Rb$_{2}$AgAsCl$_{6}$, Rb$_{2}$AgBiBr$_{6}$, Rb$_{2}$AgSbCl$_{6}$, K$_{2}$AgBiBr$_{6}$ and Cs$_{2}$AgAsCl$_{6}$). These compounds deserve to be studied further and further calculations related to their absorption spectra and intrinsic stability can be carried out in the future. 

\section*{Discussion}

From the above results, we see that unsupervised learning using agglomerative hierarchical clustering can be a powerful tool for materials discovery especially on small sized datasets where most other machine learning algorithms fail to achieve decent accuracy. This approach can be generalized to a large number of materials science problems owing to the fairly common issue of small sized datasets. This method will also save efforts in feature engineering as other machine learning algorithms based on regression trees, neural networks, etc. need to have a higher-dimensional feature vector for improved performance. This method is easily extendable to many other materials science problems with small dataset constraints. Through our DFT studies, we show that it is possible to further screen out a few compounds which show the most promising properties for photovoltaic applications and these compounds deserve to be studied by experimental synthesis and more advanced computational techniques. Consequently, this formulation of a hierarchical clustering based framework is expected to minimize the efforts in the computational screening of feasible molecules in a boundless search space.

\section*{Methods}

\subsection*{Feature engineering}

At first, 32 features were selected to construct a 32-dimensional feature vector. The features are based on correlation with band gaps according to previous reports. The dataset is in a csv file format initially which is converted into a pandas dataframe. For the data to be used in a suitable format for the AHC algorithm input, it is converted into an array comprising sub-arrays of 32 features corresponding to each compound.

\subsection*{AHC Algorithm}

The agglomerative hierarchical clustering is performed using the Scipy library in Python \cite{virtanen2020scipy}. The threshold for the division of clusters is chosen in such a way that we get 10 clusters. The AHC algorithm gives us the advantage of adjusting the partition line to get any number of clusters. This implies that the number of clusters is dynamic and can be modified at any time depending on our task. We choose 10 clusters in our study as the algorithm performs well with well-differentiated clusters. In the AHC algorithm, the similarity measure used to calculate the similarity between samples is the Euclidean distance. The cluster dissimilarity was measured using the ward linkage \cite{ward1963hierarchical}. Nodes are formed when each sample is reconnected step by step. A bottom-up tree diagram hierarchy known as dendrogram is implemented to organize the nodes. A single sample is represented by the leaf nodes of a tree and non-leaf nodes are generally obtained by merging similar or close sample sets. 

\subsection*{Ab initio Calculations}

All the calculations are performed using the Vienna ab initio simulation package
(VASP) in the framework of density functional theory (DFT). For the calculations, we
make use of the all-electron-like projector augmented wave (PAW)
pseudopotentials and the Perdew-Burke-Ernzerhof (PBE) exchange correlation
potential as implemented in the VASP code. The pseudopotentials are used with an
energy cutoff of 500 eV for the plane-wave basis functions. The k-point mesh that
is used for structure relaxation is the Monkhorst-Pack k-point mesh of 8 x 8 x 8. For
the calculation of bandstructures, the Monkhorst-Pack
k-point mesh is further increased to 12 x 12 x 12. The cell is completely optimized
including the lattice vectors and atomic positions. We make use of the criterion
that the calculated force on each atom is smaller than 0.01 eV/Å during the
structure relaxation calculations. To relax the structural parameters, the
generalized gradient approximation (GGA) of Perdew– Burke–Ernzerhof (PBE) is
employed. For the calculation of electronic properties, the hybrid nonlocal
exchange-correlation functional (HSE06) is used as it tends to give more accurate
results for the prediction of band gaps (GGA tends to underestimate the band gaps).
The spin-orbit coupling (SOC) effect is also taken into account because of heavy
valence electrons from the B'-site cations.

\section*{Data Availability}

Data for training the model is available upon request.

\section*{Code Availability}
The hierarchical clustering model codes employed in this work are available at: https://github.com//scipy//scipy.

\section*{Author Information}

\subsection*{Affiliations}
\textit{Department of Physics, Indian Institute of Technology Roorkee, Roorkee, Uttarakhand, 247667, India} \\
Utkarsh Saha \& Soumitra Satapathi \\
\\
\textit{Theoretical Sciences Unit, Jawaharlal Nehru Centre for Advanced Scientific Research, Bangalore, Karnataka, 560064, India} \\
Koyendrila Debnath

\section*{Competing interests}

The authors declare no competing interests.

\bibliographystyle{ieeetr}
\bibliography{citation}

\newpage

\section*{Supplementary Information}
\begin{table}[ht]
\caption{List of elemental properties chosen for each atom for feature set \\ preparation} 
\medskip
\centering 
\begin{tabular}{c c c} 
\hline\hline 
Elemental property & Description & Unit \\ [0.5ex] 
\hline 
$\chi$ & Pauling's electronegativity & eV \\ 
E$_{ip}$ & Ionization potential & eV \\
h & Highest occupied atomic level & eV\\
l & Lowest unoccupied atomic level & eV \\
r$_{s}$ & s-valence orbital radius of isolated neutral atom & {\AA} \\
r$_{p}$ & p-valence orbital radius of isolated neutral atom & {\AA} \\
r$_{d}$ & d-valence orbital radius of isolated neutral atom & {\AA} \\
D & Atomic distance between cation and the nearest halogen atom & {\AA} \\ [1ex]
\hline 
\end{tabular}
\label{table:nonlin} 
\end{table}

\begin{table}[hbt!]
\caption{List of compounds with their band gaps in Cluster 7} 
\medskip
\centering 
\begin{tabular}{c c c} 
\hline\hline 
Compound & Space Group & Band Gap (eV) \\ [0.5ex] 
\hline 
Cs$_{2}$AgBiBr$_{6}$ & Cubic & 1.1399 \\ 
Cs$_{2}$CuBiBr$_{6}$ & Cubic & 0.5649 \\
K$_{2}$CuBiBr$_{6}$ & Cubic & 0.5536 \\
Rb$_{2}$CuBiBr$_{6}$ & Cubic & 0.5582 \\
K$_{2}$AgBiBr$_{6}$ & Cubic & 1.1002 \\
Rb$_{2}$AgBiBr$_{6}$ & Cubic & 1.1156 \\
Cs$_{2}$AgBiBr$_{6}$ & Ortho & 1.1666 \\
Cs$_{2}$CuBiBr$_{6}$ & Ortho & 0.5429 \\ 
K$_{2}$AgBiBr$_{6}$ & Ortho & 1.4152 \\
Rb$_{2}$AgBiBr$_{6}$ & Ortho & 1.3147 \\
K$_{2}$CuBiBr$_{6}$ & Ortho & 0.8057 \\
Rb$_{2}$CuBiBr$_{6}$ & Ortho & 0.6378 \\
K$_{2}$CuBiCl$_{6}$ & Ortho & 1.1863 \\
Rb$_{2}$CuBiCl$_{6}$ & Ortho & 1.3267 \\
K$_{2}$AgBiCl$_{6}$ & Ortho & 1.7561 \\
Rb$_{2}$AgBiCl$_{6}$ & Ortho & 1.6589 \\
Cs$_{2}$AgBiCl$_{6}$ & Ortho & 1.5403 \\
Cs$_{2}$CuBiCl$_{6}$ & Ortho & 0.6710 \\
Cs$_{2}$CuBiCl$_{6}$ & Cubic & 0.7609 \\
K$_{2}$CuBiCl$_{6}$ & Cubic & 0.7442 \\
Rb$_{2}$CuBiCl$_{6}$ & Cubic & 0.7525 \\
Cs$_{2}$AgBiCl$_{6}$ & Cubic & 1.5671 \\
K$_{2}$AgBiCl$_{6}$ & Cubic & 1.5173 \\
[1ex]
\hline 
\end{tabular}
\label{table:nonlin} 
\end{table}

\newpage

\begin{table}[hbt!]
\medskip
\centering 
\begin{tabular}{c c c} 
\hline\hline 
Compound & Space Group & Band Gap (eV) \\ [0.5ex] 
\hline 
Rb$_{2}$AgBiCl$_{6}$ & Cubic & 1.5347 \\
Cs$_{2}$AgSbCl$_{6}$ & Cubic & 1.2481 \\
Cs$_{2}$CuSbCl$_{6}$ & Cubic & 0.6478 \\
K$_{2}$CuSbCl$_{6}$ & Cubic & 0.6278 \\]
Rb$_{2}$CuSbCl$_{6}$ & Cubic & 0.6353 \\
K$_{2}$AgSbCl$_{6}$ & Cubic & 1.2770 \\
Rb$_{2}$AgSbCl$_{6}$ & Cubic & 1.3030 \\
Cs$_{2}$AgAsCl$_{6}$ & Cubic & 1.2792 \\
Cs$_{2}$CuAsCl$_{6}$ & Cubic & 0.5367 \\
K$_{2}$CuAsCl$_{6}$ & Cubic & 0.5018 \\
Rb$_{2}$CuAsCl$_{6}$ & Cubic & 0.5151 \\
K$_{2}$AgAsCl$_{6}$ & Cubic & 1.1805 \\
Rb$_{2}$AgAsCl$_{6}$ & Cubic & 1.2196 \\
Cs$_{2}$CuSbCl$_{6}$ & Ortho & 0.5323 \\
Rb$_{2}$CuSbCl$_{6}$ & Ortho & 0.6550 \\
K$_{2}$CuSbCl$_{6}$ & Ortho & 1.2445 \\
Cs$_{2}$AgSbCl$_{6}$ & Ortho & 1.3456 \\
Rb$_{2}$AgSbCl$_{6}$ & Ortho & 1.5490 \\
K$_{2}$AgSbCl$_{6}$ & Ortho & 1.7291 \\
Cs$_{2}$AgAsCl$_{6}$ & Ortho & 1.2717 \\
K$_{2}$AgAsCl$_{6}$ & Ortho & 1.6011 \\
Rb$_{2}$AgAsCl$_{6}$ & Ortho & 1.3779 \\
Cs$_{2}$CuAsCl$_{6}$ & Ortho & 0.3624 \\
K$_{2}$CuAsCl$_{6}$ & Ortho & 0.6427 \\
Rb$_{2}$CuAsCl$_{6}$ & Ortho & 0.5129 \\
Cs$_{2}$AgSbBr$_{6}$ & Cubic & 0.8519 \\
Cs$_{2}$CuSbBr$_{6}$ & Cubic & 0.3254 \\
K$_{2}$CuSbBr$_{6}$ & Cubic & 0.2896 \\
Rb$_{2}$CuSbBr$_{6}$ & Cubic & 0.3028 \\
K$_{2}$AgSbBr$_{6}$ & Cubic & 0.7905 \\
Rb$_{2}$AgSbBr$_{6}$ & Cubic & 0.8142 \\
Cs$_{2}$AgAsBr$_{6}$ & Cubic & 0.7810 \\
K$_{2}$AgAsBr$_{6}$ & Cubic & 0.7009 \\
Rb$_{2}$AgAsBr$_{6}$ & Cubic & 0.7293 \\
Cs$_{2}$CuAsBr$_{6}$ & Cubic & 0.2252 \\
K$_{2}$CuAsBr$_{6}$ & Cubic & 0.1748 \\
Rb$_{2}$CuAsBr$_{6}$ & Cubic & 0.1950 \\
Cs$_{2}$CuAsBr$_{6}$ & Ortho & 0.2010 \\
K$_{2}$CuAsBr$_{6}$ & Ortho & 0.5174 \\
Rb$_{2}$CuAsBr$_{6}$ & Ortho & 0.2772 \\
Cs$_{2}$AgAsBr$_{6}$ & Ortho & 0.7789 \\
K$_{2}$AgAsBr$_{6}$ & Ortho & 1.2224 \\
Rb$_{2}$AgAsBr$_{6}$ & Ortho & 1.0250 \\
Cs$_{2}$AgSbBr$_{6}$ & Ortho & 0.8555 \\
K$_{2}$AgSbBr$_{6}$ & Ortho & 1.3449 \\
Rb$_{2}$AgSbBr$_{6}$ & Ortho & 1.1724 \\
Cs$_{2}$CuSbBr$_{6}$ & Ortho & 0.3218 \\
K$_{2}$CuSbBr$_{6}$ & Ortho & 0.6292\\
Rb$_{2}$CuSbBr$_{6}$ & Ortho & 0.4518 \\
[1ex]
\hline 
\end{tabular}
\label{table:nonlin} 
\end{table}

\end{document}